\documentclass[twocolumn, twocolappendix]{aastex63}
\usepackage{amsmath,amstext}
\usepackage[T1]{fontenc}
\usepackage{apjfonts} 
\usepackage[figure,figure*]{hypcap}
\usepackage{commath, mathrsfs}
\usepackage{color, upgreek}

\usepackage[version=4]{mhchem}
 
\newcommand{\eV}{\,{\rm eV}}
\renewcommand{\sec}{{\,\rm s}}

\newcommand{\AU}{ \, {\rm au}}
\newcommand{\au}{\AU}
\newcommand{\Msun}{{\rm \, M_{\odot}}}
\newcommand{\Mearth}{{\rm \, M_{\oplus}}}

\newcommand{\yr}{\, {\rm yr} }

\newcommand{\Myr}{\, {\rm Myr} }

\newcommand{\dd}{{\rm d}}

\newcommand{\cm}{{\, \rm cm}}

\newcommand{\K}{{\rm K}}
\newcommand{\erg}{{\rm \, erg}}

\ifx\arcsec\undefined
\newcommand{\arcsec}{\, {\rm arcsec}}
\fi

\renewcommand{\arraystretch}{1.4}

\newcommand{\eq}[1]{\begin{equation}#1\end{equation}}

\newcommand{\braket}[1]{\left(#1\right)}

\newcommand{\e}[1]{\times 10^{#1}}

\newcommand{\beqas}{\begin{eqnarray*}}
\newcommand{\eeqas}{\end{eqnarray*}}

\ifx\del\undefined
\newcommand{\del}[2]{\frac{\dd #1}{\dd #2}}
\fi

\ifx\appref\undefined
\newcommand{\appref}[1]{Appendix~\ref{#1}}
\fi
\newcommand{\fref}[1]{Figure~\ref{#1}}

\ifx\secref\undefined
\newcommand{\secref}[1]{Section~\ref{#1}}
\fi
\newcommand{\eqnref}[1]{Eq.(\ref{#1})}

\newcommand{\mum}{\, {\rm \mu m}}

\newcommand{\cs}{c_{\rm s}}

\ifx\dif\undefined
\newcommand{\dif}[2]{\frac{{\rm d}#1}{{\rm d}#2}}
\fi

\makeatletter
\newcommand*{\rom}[1]{\expandafter\@slowromancap\romannumeral #1@}
\makeatother
\ifx\ion\undefined
\newcommand\ion[2]{#1$\;${\small\rmfamily\rom{#2}}\relax}
\fi

\newcommand{\Kelvin}{{\rm \, K}}

\newcommand{\LEUV}{\Phi_{\rm EUV}}

\ifx\fh\undefined
\newcommand{\fh}{f_{\rm h}}
\fi

\ifx\rmxaa\undefined
\newcommand{\rmxaa}{}
\fi

   \newcommand{\mph}{\dot{M}_{\rm ph}}
   
	\renewcommand{\mum}{{\rm \,\upmu m}}
	\newcommand{\um}{\mum}
	
	\renewcommand{\Msun}{{\,M_\odot}}
	
	\renewcommand{\Mearth}{{\, M_\oplus}}
	\newcommand{\Mdisk}{M_\mathrm{disk}}

    \newcommand{\Mdot}[1]{\dot{M}_\mathrm{#1}}
    \newcommand{\tlife}{t_\mathrm{life,g}}

\shorttitle{Gas-Rich Debris Disks Around Intermediate-Mass Stars}
\shortauthors{Nakatani et al.}

\begin{document}

\title{A Primordial Origin for the Gas-Rich Debris Disks Around Intermediate-Mass Stars}
\author[0000-0002-1803-0203]{Riouhei Nakatani}
\affiliation{NASA Jet Propulsion Laboratory, California Institute of Technology, 4800 Oak Grove Dr, Pasadena, CA 91109, USA}
\affiliation{RIKEN Cluster for Pioneering Research, 2-1 Hirosawa, Wako-shi, Saitama 351-0198, Japan}
\email{ryohei.nakatani@jpl.nasa.gov}
\author[0000-0001-8292-1943]{Neal J. Turner}
\affiliation{NASA Jet Propulsion Laboratory, California Institute of Technology, 4800 Oak Grove Dr, Pasadena, CA 91109, USA}
\author[0000-0000-0000-0000]{Yasuhiro Hasegawa}
\affiliation{NASA Jet Propulsion Laboratory, California Institute of Technology, 4800 Oak Grove Dr, Pasadena, CA 91109, USA}
\author[0000-0002-2700-9676]{Gianni Cataldi}
\affiliation{National Astronomical Observatory of Japan, Osawa 2-21-1, Mitaka, Tokyo 181-8588, Japan}
\author[0000-0000-0000-0000]{Yuri Aikawa}
\affiliation{Department of Astronomy, Graduate School of Science, The University of Tokyo, Tokyo 113-0033, Japan}
\author[0000-0000-0000-0000]{Sebasti\'an Marino}
\affiliation{Department of Physics and Astronomy, University of Exeter, Stocker Road, Exeter, EX4 4QL, UK}
\author[0000-0000-0000-0000]{Hiroshi Kobayashi}
\affiliation{Department of Physics, Nagoya University, Furo-cho, Chikusa-ku, Nagoya, Aichi 464-8602, Japan}
% \author[0000-0003-2309-8963]{Rolf Kuiper}
% \affiliation{Institute of Astronomy and Astrophysics, University of T\"ubingen, Auf der Morgenstelle 10, D-72076 T\"ubingen, Germany}
% \author[0000-0002-7058-7682]{Hideko Nomura}
% \affiliation{National Astronomical Observatory Japan (NAOJ), Osawa 2-21-1, Mitaka, Tokyo 181-8588, Japan}
% \author[0000-0003-3283-6884]{Yuri Aikawa}
% \affiliation{Department of Astronomy, School of Science, The University of Tokyo, 7-3-1 Hongo, Bunkyo, Tokyo 113-0033, Japan}

\begin{abstract}
While most debris disks consist of dust with little or no gas, a fraction has significant amounts of gas detected via emission lines of CO, ionized carbon, and/or atomic oxygen.  Almost all such gaseous debris disks known are around A-type stars with ages up to 50~Myr.  We show, using semi-analytic disk evolution modeling, that this can be understood if the gaseous debris disks are remnant protoplanetary disks that have become depleted of small grains compared to the interstellar medium.  Photoelectric heating by the A-stars' FUV radiation is then inefficient, while the stars' EUV and X-ray emissions are weak owing to a lack of surface convective zones capable of driving magnetic activity.  In this picture, stars outside the range of spectral types from A through early B are relatively hard to have such long-lived gas disks.  Less-massive stars have stronger magnetic activity in the chromosphere, transition region, and corona with resulting EUV and X-ray emission, while more-massive stars have photospheres hot enough to produce strong EUV radiation.  In both cases, primordial disk gas is likely to photoevaporate well before 50~Myr.  These results come from {0D} disk evolution models where we incorporate internal accretion stresses, MHD winds, and photoevaporation by EUV and X-ray photons with luminosities that are functions of the stellar mass and age.  A key issue this work leaves open is how some disks become depleted in small dust so that FUV photoevaporation slows.  Candidates include grains' growth, settling, radial drift, radiation force, and incorporation into planetary systems.
\end{abstract}

\keywords{}

%------------------------- introduction ---------------------
\section{Introduction}  \label{sec:introduction}
    A fundamental goal of planet formation research is to learn the evolutionary pathways that protoplanetary disks (PPDs) follow to become planetary systems and their associated debris disks.
    Classically the PPD lifetime is estimated to be no longer than $10 \Myr$, based on near- and mid-infrared (IR) observations which trace the inner hot dust \citep[$\lesssim 1\text{--}10\au$;][]{2001_Haisch, 2007_Meyer, 2007_Hernandez, 2009_Mamajek, 2014_Ribas, 2015_Ribas} and ultraviolet (UV) and H$\alpha$ observations tracing accreting gas \citep[e.g.,][]{2009_KennedyKenyon, 2010_Fedele, 2010_Sicilia-Aguilar}.  Ten Myr is thus
    considered a typical timescale on which gas-rich PPDs evolve into gas-free debris disks. 
    
    An explanation is therefore needed for the \replaced{several tens of}{more than 20} gaseous debris disks observed around stars older than $10 \Myr$ \citep[e.g.,][]{2013_Kospal, 2016_Marino, 2016_Lieman, 2017_Moor, 2017_Hughes,  2017_Higuchi, 2019_Higuchia, 2019_Higuchib, 2020_Kral}.
    Two hypotheses have been proposed for the origins of this gas.
    In the primordial-origin scenario, the gas is a remnant of the PPD. 
    In the secondary-origin scenario, the gas was released more recently from volatiles in planetesimals associated with the dust in the debris disks. 

    While the secondary-origin scenario has been extensively examined through modeling \citep{2016_Kral, 2019_Kral, 2019_Moor, 2019_Matra, 2020_Marino, 2020_Cataldi, 2022_Marino}, 
    less attention has been paid to the primordial-origin scenario \citep{2021_Nakatani, 2022_Smirnov-Pinchukov, 2023_Iwasaki}.
    A primordial origin has been considered unlikely because of the short canonical PPD gas disk lifetime. 
    However, hydrodynamical modeling by \cite{2021_Nakatani} shows that if the PPD around an intermediate-mass star ($M_* \approx 2\Msun$) is depleted in submicron grains, including polycyclic aromatic hydrocarbons (PAHs), then the photoelectric heating produced when FUV photons strike grains is weak enough to slow photoevaporation so the gas could survive well past $10 \Myr$ and into the debris disk stage
    {--- without small-grain depletion, the gas lifetimes are most likely $\lesssim 10\Myr$ regardless of stellar masses \citep{2023_Komaki}.}
    Primordial origins thus have the potential to account for the occurrence of gaseous debris disks around A stars \citep{2016_Lieman,2017_Moor, 2018_Hughes}. % more easily than has been possible under secondary-origin models \citep{2019_Matra, 2020_Marino}.  
    % A further reason to explore primordial origins is that this picture offers an explanation for some of the disks' observed spatially extended gas compared to the dust. %spatial offsets between the dust and the gas. 

    The outstanding questions include what gas lifetimes to expect quantitatively in the primordial origin scenario, how these depend on the stellar mass, % and amount of dust remaining in the gas, 
    and whether the results are consistent with the observed occurrence rates of gaseous debris disks. 
    We here address these issues by constructing disk evolution models using a semi-analytic, {0D} approach. 
\deleted{    We outline the main processes acting on the gas in \secref{sec:overall_picture}, derive the lifetimes of small-grain-depleted disks for various stellar masses in \secref{sec:lifetime_estimate}, and compare them with the detection rate of gaseous debris disks in \secref{sec:observational_disks}.  The results are summarized in \secref{sec:conclusion}.}

\section{Overall picture of gas disk evolution}  \label{sec:overall_picture}
    Theoretical works indicate the gas disks are dispersed by
    accretion onto the central star \citep{1973_ShakuraSunyaev, 1974_LindenbellPringle} and by ejection in winds, launched either by
    magnetic forces \citep[e.g.,][]{2009_SuzukiInutsuka, 2013_Bai_a, 2013_Bai_b}
    or by photoevaporation \citep[e.g.,][]{1993_Shu_b, 1994_Hollenbach, 1998_Richling}.
    These three processes' relative importance varies over the course of disk evolution: 
    accretion and/or magnetohydrodynamic (MHD) winds dominate the mass loss in the early stage when the disk is relatively massive,
    while photoevaporation dominates later when the disk mass decreases to $\lesssim 10^{-2}\Msun$ at ages of 1--10$\Myr$ \cite[e.g.,][]{2001_Clarke, 2015_Gorti, 2017_Carrera, 2020_Kunitomo, 2021_Kunitomo, 2023_Weder, 2023_Komaki}. 

    This picture of time-varying disk dispersal processes is consistent with several observational facts. 
    For instance, class~0/I jets have much larger mass-loss rates ($10^{-7}\text{--}10^{-6}\Msun\yr^{-1}$) than photoevaporation can provide, implying the ejection is driven by MHD effects in these early stages. 
    For another example, the luminosity of a photoevaporative wind tracer, the [\ion{Ne}{2}]~12.8$\um$ low-velocity component, is higher for more evolved disks, i.e., slow accretors ($\sim 10^{-8}\Msun \yr^{-1}$) and inner-dust-depleted disks \citep{2020_Pascucci}. 
    
    Jet and wind mass-loss rates in the upper end of the range imply possibly enough material crossing the lines of sight between star and disk to block much of the stellar ultraviolet (UV) and X-ray radiation. 
    This can slow or stop the disk's photoevaporation \citep[for a recent review, see][]{2022_Pascucci}. 
    In 3D MHD modeling by \cite{2022_Takasao}, the outflow from the disk around a solar-mass star has a column density high enough to attenuate the UV and X-ray photons at accretion rates $\gtrsim 10^{-8}\Msun\yr^{-1}$. 
    Overall, a fairly advanced age and low rates of accretion and outflow appear necessary before photoevaporation can dominate gas disk dispersal.

\section{Gas Lifetime Estimate}  \label{sec:lifetime_estimate}
    {For the primordial-origin hypothesis to be a possible scenario, one of the critical necessary conditions PPDs must meet is to retain a mass reservoir beyond $10\Myr$, especially at outer radii ($\gtrsim1\text{--}10\au$) where gaseous debris disks are detected. This study's primary objective is to demonstrate its feasibility, which was previously considered implausible. Our model will be tailored for this purpose, as will described below.}
    
    % We aim to test the plausibility of the primordial-origin scenario, 
    % and therefore 
    We focus on the disks where PAH and small grains (${\lesssim} 0.01\um$) are depleted compared to the interstellar medium. 
    These are the systems %where gas potentially survives for $>10\Myr$ 
    {that potentially meet the necessary condition;}
    % since the slower grain photoelectric heating weakens the photoevaporation \citep{2021_Nakatani}. 
    % By contrast, 
    otherwise, the disks dissipating through rapid photoelectric heating most likely have short lifetimes $< 10\Myr$ \citep{2009_Gorti, 2015_Gorti, 2021_Komaki, 2023_Komaki}; we ignore this population.
    Thus, we here derive the lifetime for the sub-population of PPDs where the gas is the longest-lived.
    % and test the primordial-origin scenario by examining how the gas disk lifetime changes as a function of the stellar mass ($M_*$).
    The estimated lifetime is distinct from the average gas lifetime, which depends on the still-unknown probability of achieving a small-grain-depleted state. 
    % Deriving the lifetimes of the gas disks that survive longest at a given stellar mass enables estimating the incidence of primordial-gas-rich debris disks and thus testing the primordial-origin scenario. 
    % i.e., if we select only the disks surviving for > 10 Myr, that population should be the grain-depleted disks on the theory side, and it should be gas-rich debris disks on the observation side. 
    % we can directly test whether disk lifetimes indeed exceed $10\Myr$ under the optimal situation for gas survival; if they could not, we could conclude that the primordial-origin scenario is unlikely from the viewpoint of disk dispersal. 
    
    We do not specify here the processes responsible for grain depletion. 
    Instead, we simply assume the disks reach a small-grain-depleted state before photoevaporation begins to dominate disk dispersal, regardless of the stellar mass.
    Several effects can deplete the dust, especially for intermediate-mass stars:
    Grain growth and radial drift reduce the local dust-to-gas ratio \citep[e.g.,][]{2020_Sellek}.
    The drift is likely especially rapid around young intermediate-mass stars \citep{2022_Pinilla}. 
    The intense radiation force from intermediate-mass stars can also blow out small dust \citep{2019_Owen, 2021_Nakatani}.
    
    {
    The specific level of small-grain depletion needed to inhibit FUV photoevaporation remains uncertain. 
    \citet{2018_Nakatani, 2018_Nakatanib} explored FUV photoevaporation varying disk metallicity
    and demonstrated that strong FUV photoevaporation demands small grain abundance exceeding $\sim 0.1\text{--}1\%$ of the ISM level. 
    However, these investigations focus on a $0.5\Msun$ star with extremely high FUV luminosity, lowering oxygen and carbon abundances with metallicity, which limits the results' general applicability. 
    The critical small-grain abundance depends on stellar mass and FUV/X-ray luminosities, yet prior studies explored a limited parameter space. 
    Further investigations are warranted.
    Nevertheless, it is likely that critical small-grain abundances exist, below which FUV photoevaporation becomes negligible.
    % An estimate places this threshold around $\sim 1\%$ of the ISM level
    We anticipate this threshold to be around $\sim 0.1\text{--}1\%$ of the ISM level, following \citet{2018_Nakatani, 2018_Nakatanib}.
    }
    
    Observations suggest the PAH abundance in PPDs is at least ten times less than the interstellar value \citep[e.g.,][]{2007_Geers, 2010_Oliveira, 2013_Vicente}.  
    The hybrid disk HD~141569A is depleted to a level where photoelectric heating may drive only weak photoevaporation \citep{2003_Li, 2014_Thi}.
    { 
    Non-detection of PAHs occur with a reasonable fraction in Herbig disks, where the detection rate is $\sim 70\%$ \citep[e.g.,][]{2010_Acke}, or in T Tauri disks, where it is $\lesssim 10\%$; \citep[e.g.,][]{2006_Geers}.
    Such non-detection hints at abundances potentially over 100 times lower than interstellar levels.
    This significant depletion in the disk atmosphere is also indicated from mid-IR spectra \citep[e.g,][]{2006_Furlan, 2011_Furlan}
    and implied from IR scattered-light imaging \citep[e.g.,][]{2013_Mulders}.
    These findings lead us to infer that these disks might be in the small-grain-depleted state.
    }

\subsection{Semi-Analytic Model}
    % \subsection{Lifetimes without MHD winds}
    % We quantify the $M_*$ dependence of the gas disk lifetime, $\tlife$, following the UV-switch model \citep[][detailed in Appendix~\ref{sec:uv_switch}]{2001_Clarke}. 
    {Our model employs a 0D approach, following the disk's mass decrease from accretion and wind mass loss in the early stage and from photoevaporation in the late stage.
    It is based on the analytic lifetime estimate from} the UV-switch model \citep[][detailed in Appendix~\ref{sec:uv_switch}]{2001_Clarke}.
    In their model, the major dispersal process is viscous accretion at the early stage and switches to photoevaporation beginning at the ``switching time'' $t_0$,
    {which defines the transition point into the late stage.} 
    Thereafter, it disperses the remaining disk mass of $M_{\rm disk, 0}^\prime$. 
    This model returns $t_0$, $M_\mathrm{disk, 0}^\prime$, and {the gas disk lifetime} $\tlife$ analytically (cf.\ Eqs.\ref{eq:switching_time}--\ref{eq:evaporation_time}) once the disk mass $\Mdisk(t)$ (cf.\ \eqnref{eq:mdot_ana}) and photoevaporation rate $\mph(t)$ are given as functions of time. 

    The original UV-switch model uses purely viscous disks, but here we consider disks evolving under angular momentum and mass extraction through a magnetically-driven wind alongside the internal angular momentum redistribution by turbulent stresses.  
    Observational evidence that PPDs appear to evolve by driving magnetized winds is discussed by \citet{2017_Rafikov}, \citet{2022_Manara}, and \citet{2022_Pascucci}.
    We adopt the analytic disk evolution model of \cite{2019_Chambers} for $\Mdisk(t)$, as set out in \secref{sec:chambers}. 
    We first derive $t_0$ and $M_\mathrm{disk, 0}^\prime$ by finding the time when the disk mass loss $\dot{M}_\mathrm{disk}(t)$ and photoevaporation rates $\mph(t)$ equal;
    we will define $\mph(t)$ later.
    % The gas disk lifetime 
    $t_\mathrm{life,g }$ is then the time till the remaining mass is photoevaporated, found by numerically integrating
    \[
        \Mdisk (t = t_0) = M_{\rm disk,0}^\prime  = \int_{t_0}^{\tlife} \dd t \, \mph (t).
    \]
    The resulting $\tlife$ is insensitive to the choice of the threshold mass to define ``dispersal'', as $\Mdisk$ falls rapidly after $t = t_0$. 
    The disk's initial mass is taken as proportional to the stellar mass with $M_\mathrm{disk,0} = 0.1 M_*$ \citep[e.g.,][however a steeper scaling $\propto M_*^{1.3\text{--}2.0}$ is found by \citet{2016_Pascucci, 2016_Ansdell, 2017_Ansdell}; see \citet{2022_Manara} for a recent review]{2011_WiiliamsCieza, 2013_Andrews}. 
    % We also assume the viscous accretion flow timescale $t_\nu \propto M_*^{-1}$ so the steady-state accretion rate $\Mdot{acc,0} \propto M_\mathrm{disk, 0} t_\nu^{-1} \propto M_*^2$, as observed \citep[e.g.,][]{2004_Calvet, 2003_Muzerolle, 2005_Muzerolle, 2005_Mohanty, 2006_Natta, 2014_Alcala, 2017_Alcala, 2016_Manara, 2022_Manara}. 
    We explore $\tlife$'s dependence  on $M_\mathrm{disk,0}$ and the scaling in Appendix~\ref{sec:parameter_dependences}.

    Photoevaporation is generally driven by far-ultraviolet (FUV), extreme-ultraviolet (EUV), and X-ray radiation. 
    However, 
    \replaced{since we are interested in long-lived disks,}{as stated above,}
    we here consider only cases where FUV grain photoelectric heating is negligible, and the photoevaporative winds are driven solely by the EUV and X-ray heating associated with photoionization. % of (mainly) hydrogen and helium. %(Paths~2-(i) and (ii) in Paper~I). 
    We also assume that the disks are isolated from massive stars, so external photoevaporation is negligible.
    % {as no gaseous debris disks have been detected nearby massive stars}. 
    We use the EUV photoevaporation rate estimate 
    \begin{equation}
        \Mdot{EUV} = 4.1\e{-10}\Msun\yr^{-1} 
        \braket{\frac{\LEUV}{10^{41}\sec^{-1}}}^{1/2}
        \braket{\frac{M_*}{1\Msun}}^{1/2}
        \label{eq:mdot_euv}
    \end{equation}
    \citep{1994_Hollenbach, 2001_Clarke} since it agrees well with the mass-loss rates derived by radiation hydrodynamics simulations \citep{2021_Nakatani}. 
    Based on 1+1-D modeling, EUV photoevaporation rates were thought to increase about tenfold after the disk center is cleared out, letting the stellar EUV field dominate over the diffuse EUV from recombining ions. 
    However, 2D axisymmetric radiative transfer showed that the direct stellar component dominates even in disks without central cavities \citep{2013_Tanaka}.  Additionally, radiation hydrodynamics modeling found no significant difference in photoevaporation rates with and without cavities of various sizes \citep{2010_Owen, 2019_Picogna, 2021_Nakatani}.
    Hence, we apply the \eqnref{eq:mdot_euv} photoevaporation rates uniformly throughout disk evolution. %before and after $t_0$.
    
    The stellar EUV emission rate $\LEUV$ can be decomposed into photospheric, chromospheric ($\sim 10^4\Kelvin$), transition regional ($\sim 10^5\Kelvin$), and coronal ($\sim 10^6\Kelvin$) components. 
    We shall refer to the last three jointly as magnetic-origin EUV, \added{$\Phi_\mathrm{EUV,mag}$}.
    For the photospheric EUV component, we use \cite{2021_Kunitomo} Table~1 results from long-term stellar evolution calculations for $0.5\Msun \leq M_* \leq 5\Msun$. 
    % We take the time-dependent photospheric EUV emission rate $\Phi_\mathrm{EUV, ph}$ from the table. 
    We estimate the magnetic-origin EUV emission rate from the X-ray luminosity $L_\mathrm{X}$ using the $\LEUV\text{--}L_\mathrm{X}$ relation of \citet{2021_ShodaTakasao},
    \begin{equation}
        \log \braket{\frac{\Phi_\mathrm{EUV, mag}}{1\sec^{-1}}} = 20.40 + 0.66 \log \braket{\frac{L_\mathrm{X}}{1\erg\sec^{-1}}}.  \label{eq:Phi_EUV_mag}
    \end{equation}
    Strictly, this relation applies to solar-type stars, but we apply it to all the stars considered here ($0.5\Msun \leq M_* \leq 5\Msun$).
    Since the theoretical and observational uncertainties in \replaced{the magnetic-origin EUV emission}{$\Phi_\mathrm{EUV,mag}$} are large, we believe this estimate is a reasonable first-order approximation.
    % The time-dependent $L_\mathrm{X}$ in \cite{2021_Kunitomo} includes a conservative floor level $L_\mathrm{X}/L_* \geq 10^{-7}$, 
    % but we let $L_\mathrm{X}$ decrease below $10^{-7}$ in \eqnref{eq:Phi_EUV_mag}, to avoid causing an artificial feature in $\Phi_\mathrm{EUV, mag}$. 

    We follow \cite{2021_Kunitomo} for $L_\mathrm{X}$ time evolution, excluding the $L_\mathrm{X}/L_* = 10^{-7}$ floor applied to intermediate-mass stars in the original paper. We expect the floor not to apply universally, given that most Herbig stars do not present X-ray emissions {detected at this level} \citep[e.g.,][]{2005_Hamaguchi,2006_Stelzer}. 
    Instead, we employ $L_\mathrm{X} = \mathrm{min}(10^{-3.13}, 5.3\e{-6}\mathrm{Ro}^{-2.7}) L_*$, where $\mathrm{Ro}$ represents the Rossby number \citep{1984_MangeneyPraderie, 1984_Noyes}
    {defined as $\mathrm{Ro} = P_\mathrm{rot}/\tau_\mathrm{conv}$ with $P_\mathrm{rot}$ and $\tau_\mathrm{conv}$ being the rotational period and the convective turnover timescale. 
    \citet{2021_Kunitomo} set $P_\mathrm{rot}$ to 3\,days and calculate $\tau_\mathrm{conv}$ from the stellar evolution model.
    % We follow their technique and extract the time evolution of $\tau_\mathrm{conv}$ from \citet{2021_Kunitomo} Table~1 to calculate $L_\mathrm{X}$.
    The rotational period can range between 1--10\,days in general, but the qualitative trend in the time evolution of $L_\mathrm{X}$ is independent of $P_\mathrm{rot}$ (see Section~5.2 of \citet{2021_Kunitomo}). Thus, the variation of $P_\mathrm{rot}$ would have a limited impact on our results.}
    % calculated using empirical rotation period and stellar evolution model (see \citet{2021_Kunitomo} for detail). 
    % Its time evolution is also extracted from \citet{2021_Kunitomo} Table~1.
    The total EUV emission rate is $\LEUV = \Phi_\mathrm{EUV, ph} + \Phi_\mathrm{EUV, mag}$.
    We omit accretion-generated EUV and X-ray here
    \deleted{since it is unclear what fraction of the accretion power is emitted in the EUV and whether those photons escape the inner accretion flow (\secref{sec:overall_picture}).}
    but address their influence on $\tlife$ in Appendix~\ref{sec:accretion_generated}.

    % We find that Lyman-Werner (LW) radiation can drive weak photoevaporation, yielding slow mass loss for stars with $2\Msun \lesssim M_* \lesssim 3\Msun$ (hereafter late intermediate-mass stars; Nakatani et al.\ in prep.). 
    % Since the luminosity dependence of the LW photoevaporation rate is unknown, 
    % we do not include the LW photoevaporation rate here. 
    % However, we find the EUV photoevaporation rate from \eqnref{eq:mdot_euv} is largely similar to the LW photoevaporation rate for the late intermediate-mass stars.
    % Thus, neglecting the LW photoevaporation is unlikely to affect the derived lifetimes significantly.
    Stellar/interstellar Lyman-Werner (LW) radiation-driven photoevaporation might be important for small-grain-depleted disks around stars with $2\Msun \lesssim M_* \lesssim 3\Msun$ (hereafter late intermediate-mass stars).
    However, our radiation hydrodynamics simulations indicate that LW photoevaporation is negligible even for late intermediate-mass stars (Nakatani et al.\ in prep.).
    Thus, we do not consider LW photoevaporation in this study. 
    The dependence of LW photoevaporation on stellar mass and luminosity is currently unknown but warrants investigation in future research.

    For X-ray photoevaporation, the mass-loss rate is still under debate. 
    Some studies obtained large mass-loss rates \deleted{comparable to the FUV photoevaporation rates} \citep{2009_Ercolano, 2010_Owen, 2019_Picogna},
    while others found %concluded X-rays yield 
    weaker mass loss \citep{2009_Gorti, 2017_Wang, 2018_Nakatanib, 2021_Komaki}. 
    The different conclusions likely originate from the adopted X-ray spectra ---
    mass-loss rates are high if the spectrum has a certain level of soft X-rays \citep[$\sim 0.1{\, \rm keV}$,][]{2009_Ercolano, 2009_Gorti, 2018_Nakatanib, 2022_Sellek}, from the incorporated cooling processes --- the studies with large X-ray photoevaporation rates neglect important cooling processes, \deleted{namely molecular and adiabatic cooling,} and from the adopted numerical methods.
    To cover the uncertainty of X-ray photoevaporation rates, we estimate the lifetimes with and without X-ray photoevaporation. 

    For X-ray photoevaporation rates, we use the formula of \cite{2012_Owen},
    \deleted{\footnote{Using the formula of \citet{2019_Picogna} instead results in very similar lifetimes.}}
    \begin{equation}
        \Mdot{X} = 4.8\e{-9}\Msun\yr^{-1}
        \braket{\frac{L_\mathrm{X}}{10^{30}\erg\sec^{-1}}}^{1.14}
        \braket{\frac{M_*}{1\Msun}}^{-0.148}.
        \label{eq:mdot_owen}
    \end{equation}
    % and the second is a recent, modified version,
    % \begin{equation}
    %     \log \braket{
    %     \frac{\Mdot{X} }{1\Msun\yr^{-1}}}
    %     =   A \exp\left[  \frac{\braket{\ln\left[\log \tilde{L}_\mathrm{X}\right] - B}^2}{C}\right] + D,
    %     \label{eq:mdot_picogna}
    % \end{equation}
    % with $(A, B, C, D) = (-2.7326, 3.3307, -2.9868\e{-3}, -7.2580)$
    % and $\tilde{L}_\mathrm{X} \equiv ({L_\mathrm{X}}/{1\erg \sec^{-1}})$ \citep{2019_Picogna}. 
    % In [140]: 10**np.exp(3.3307)
    % Out[140]: 9.076227829592288e+27
    % This second equation applies only for $L_\mathrm{X} \gtrsim 10^{28}\erg \sec^{-1}$ because of its quadratic form. 
    % We therefore use a floor value $L_\mathrm{X}/L_* = 10^{-7}$ \citep[as in][]{2021_Kunitomo} in \eqnref{eq:mdot_picogna} only.
    % This prevents $L_\mathrm{X}$ from falling below $\sim 10^{28}\erg\sec^{-1}$ in all cases we consider here.
    (Using the formula of \citet{2019_Picogna} instead results in very similar lifetimes.)
    Given the disagreement over X-ray photoevaporation rates discussed above, \eqnref{eq:mdot_owen} must be considered an upper limit, 
    so the corresponding $\tlife$ is a lower limit. 
    The total photoevaporation rate is 
    \[
        \Mdot{ph} =
        \left\{
        \begin{array}{cl}
        \Mdot{EUV} & \text{(w/o X-ray photoevaporation)} \\
        \mathrm{max}\braket{\Mdot{EUV}, \Mdot{X}} & \text{(w/ X-ray photoevaporation)} 
        \end{array}
        \right.
        .
    \]
    We calculate the lifetimes for two cases: (i) EUV only and (ii) EUV + X-ray with Eqs.\eqref{eq:mdot_euv} and \eqref{eq:mdot_owen}. 
    In Appendix~\ref{sec:X-ray_spread}, we additionally discuss how the observed spread in $L_\mathrm{X}$ \citep[e.g.,][]{2007_Gudel} can affect $\tlife$.
    {Note that using the sum instead of $\mathrm{max}\braket{\Mdot{EUV}, \Mdot{X}}$ does not make a difference, as $\Mdot{X}$ dominates $\Mdot{EUV}$ significantly.}

    \begin{figure}
        \centering
        \includegraphics[clip, width = \linewidth]{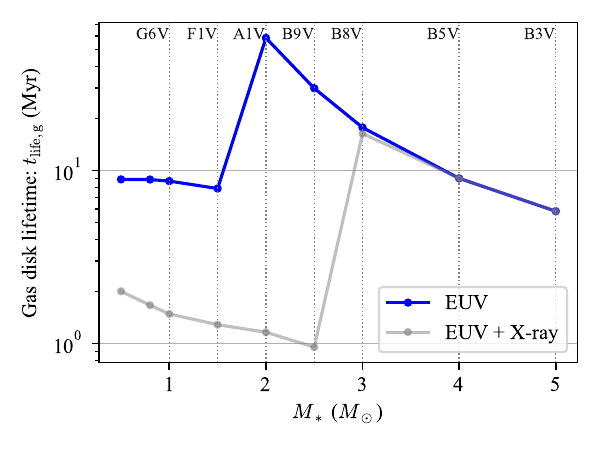}
        \caption{
        Estimated gas disk lifetime $\tlife$ as a function of $M_*$. 
        The blue line indicates $\tlife$ for the EUV-only case. 
        The gray %dashed 
        line adds X-ray photoevaporation. 
        The spectral type corresponding to each $M_*$ is presented on the top axis.
        Note that these lifetimes apply only for small-grain-depleted disks and not to disks where FUV photoelectric heating drives photoevaporation.
        % { (Right) Same as the left plot, but the dashed lines show $t_0$. For comparison, $\tlife$ is also overplotted with thinner lines. }
        % {
        % NOTE: Using Tanaka+13 equation shortens the lifetimes, although the $\tlife$--$M_*$ relations are qualitatively the same. See Database/MachineReadableTable/Kunitomo2021 for comparison.
        % }
        }
        \label{fig:t_life-Mstar}
        % python3.9 ~/Dropbox/Liouhei_Ryohei/PythonScripts/HybridDisk_paperfigs.py life 
    \end{figure}

\subsection{Results}    
    % \fref{fig:t_life-Mstar} top panel shows the estimated $\tlife$ as a function of $M_*$ when $M_\mathrm{disk,0} = 0.1M_*$ and $t_\nu = 10^5\yr (M_*/\Msun)^{-1}$. 
    {To provide an overview, we start with}
    \fref{fig:t_life-Mstar}, %shows 
    the estimated $\tlife$ as a function of $M_*$ using the \cite{2019_Chambers} model for disks dominated by slow winds.   
    Here the accretion is predominantly driven by the wind, which also extracts a significant fraction of the disk's mass (see Section~4.3 of the original paper for details).
    % As a result, the disk mass declines faster and the photoevaporation-dominated epoch begins earlier. 
    The figure shows the corresponding spectral type at each $M_*$ determined by $T_\mathrm{eff}$ when the stars reach the zero-age-main-sequence (ZMAS) in the table of \cite{2021_Kunitomo}. 
    The relation between the spectral type and $T_\mathrm{eff}$ is taken from \cite{2013_Pecaut}.
    The table of \cite{2021_Kunitomo} ends before stars with $M_* \leq 1\Msun$ reach the ZAMS. 
    Thus, for $M_* = 1\Msun$, we used the last $T_\mathrm{eff}$ value in the table to determine the spectral type; the spectral types of $M_* < 1\Msun$ stars are undetermined, so not specified in \fref{fig:t_life-Mstar}.

    % However, $\tlife$ is much shorter (compare the upper and lower panels in \fref{fig:t_life-Mstar}), consistent with observationally inferred disk lifetimes. 
    In the EUV-only case (blue), the lifetime peaks at $M_* = 2\Msun$, or equivalently spectral types around~A0. 
    Lifetimes longer than $10\Myr$ are found only around the late intermediate-mass stars ($2\Msun \lesssim M_* \lesssim 3\Msun$), corresponding to spectral types between~F0 and~B8.  
    The surface convective zones of these stars disappear at $\sim1$--$10\Myr$, 
    and the magnetic-origin EUV emission is already weak by the time photoevaporation comes to dominate mass loss. 
    The results indicate that primordial gas disks relatively easily survive to the ages of young debris disks around A- and late B-type stars. 
    
    Stars with $M_* \geq 4\Msun$ (hereafter early intermediate-mass stars) produce much stronger photospheric EUV emission owing to their high $T_\mathrm{eff}$, leading to shorter $\tlife$. 
    For stars of Solar mass and below, the magnetic-origin EUV emission remains strong throughout the disk dispersal. 

    % The gas lifetimes in \fref{fig:t_life-Mstar} top panel are too long because we have assumed the disks' accretion flows come from internal stresses alone.
    % However, the PPD population does not follow the relationships between disk parameters expected in the viscous accretion picture \citep{2017_Rafikov, 2018_Ansdell}. 
    % Furthermore, the viscosities estimated from local line widths and dust scale heights ($\alpha \lesssim 10^{-3}$) imply accretion rates lower than the values $\gtrsim 10^{-8}\Msun\yr^{-1}$ measured from hydrogen line luminosities \citep{2015_Flaherty, 2016_Pinte}. 
    % The mass accretion rate is known to be lower than estimated from H$\alpha$ emission ($\sim 10^{-8}\Msun\yr^{-1}$) as long as we adopt the viscous parameter estimated from the local line width ($\sim 10^{-3}$).  % it's not always Halpha (Brgamma is used in Pinte+16)
    % There is growing observational evidence that PPDs do not likely evolve in a purely viscous manner. 
    % Besides, turbulent viscosity is estimated to be too small to explain the observed accretion rates \citep{2016_Pinte}. % according to the dust scale height in the HL Tau disk
    % These facts imply some other process must contribute to the disks' evolution.  The leading candidate is magnetically-driven winds.
    % In the following section, we will see that incorporating the magnetized winds reduces $\tlife$. 

    X-ray photoevaporation (gray line in \fref{fig:t_life-Mstar}) reduces the lifetime by 1--2 orders of magnitude for $\lesssim 3\Msun$ compared to the EUV-only case. 
    This is because $L_\mathrm{X}$ is kept high at $\gtrsim 10^{31}\erg \sec^{-1}$, corresponding to $\Mdot{X} \sim 5\e{-8}\text{--} 10^{-7} \Msun \yr^{-1}$, until the disk completely disperses. 
    However, as mentioned above, the adopted X-ray photoevaporation rates \replaced{are likely}{might be} overestimated. \deleted{for the simplified thermal treatment and neglecting molecular and adiabatic cooling.} 
    While X-ray photoevaporation can explain observational wind diagnostics \citep[e.g.,][]{2020_Weber, 2022_Rab},
    such high X-ray photoevaporation rates are also known to be unfavorable for 
    %explaining the observed low-mass disks and accretion rates with viscous disk models \citep{2020_Sellek, 2023_Alexander} 
    explaining the observed disk mass and accretion rates with viscous disk models \citep{2020_Sellek, 2023_Alexander} and magnetized disk models \citep{2023_Weder}; smaller mass-loss rates as EUV photoevaporation alone are preferred.
    \deleted{Furthermore, our EUV+X-ray models predict inner disk lifetimes much shorter than observations, prompting us to reconsider the validity of X-ray photoevaporation treatment;
    In Appendix~\ref{sec:inner_disk_lifetime}, we present the estimated inner disk lifetimes as well as $t_0$. }
    {However, we do not rule out the effectiveness of X-rays in driving photoevaporation, as they indeed deposit energy to the gas, but with somewhat much smaller mass-loss rates than \eqnref{eq:mdot_owen}. If so, the lifetimes of the EUV models set upper limits for low-mass stars.}
    For the early intermediate-mass stars ($\geq 4\Msun$), $\tlife$ matches the EUV-only case since the convective zone disappears before $1\Myr$.

    % We also compute $\tlife$ for several values of $0.03M_* \leq M_\mathrm{disk,0} \leq 0.1 M_*$ and $0.1\Myr \leq t_\nu \leq 0.5\Myr$. 
    % The $\tlife$--$M_*$ relations resemble \fref{fig:t_life-Mstar} top panel, 
    % except that a small $M_\mathrm{disk, 0}\approx 0.03M_*$ reduces $\tlife$ at $1.5\Msun$ to $\sim 10\Myr$. 
    % In this case the convective zone persists till $\sim 10\Myr$, enabling photoevaporation to remove the entire disk.  
    % Thus, $\tlife$ is sensitive to the initial disk mass.   

    We then explore the dependence on the parameters of the MHD wind-driven accretion and mass loss across the fast-wind and laminar-disk regimes in \cite{2019_Chambers}.
    In the fast-wind case, the wind mass loss is negligible, and most of the accretion is driven by the wind, while in the laminar case, wind mass loss is substantial, and the wind drives almost all the accretion (Sections~4.2 and~4.4 of the original paper). 
    \begin{figure}
        \centering
        \includegraphics[clip, width = \linewidth]{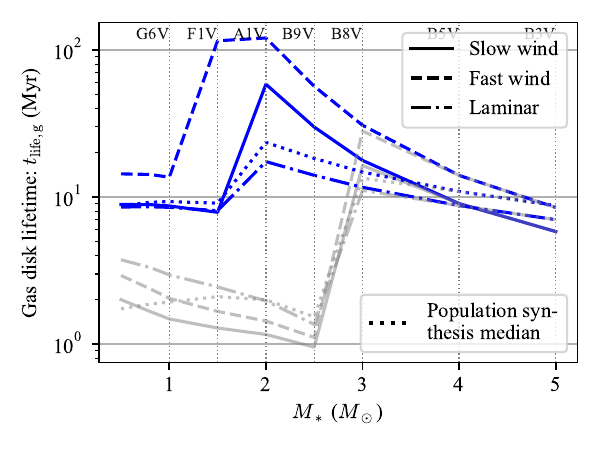}
        \caption{ Same as \fref{fig:t_life-Mstar}. Solid, dashed, and dot-dashed lines show lifetimes of slow-wind, fast-wind, and laminar disks, respectively. The dotted lines show the median lifetimes of the population synthesis in Appendix~\ref{sec:parameter_dependences}.}
        \label{fig:multi_lifetimes}
        % python3.9 ~/Dropbox/Liouhei_Ryohei/PythonScripts/HybridDisk_paperfigs.py life_multi
    \end{figure}
    The resulting $\tlife$--$M_*$ relation (\fref{fig:multi_lifetimes}) is essentially the same as \fref{fig:t_life-Mstar}, consistently peaking at $2\Msun$. 
    Small differences are that in the fast-wind model, $\tlife$ at $1.5\Msun$ becomes comparable to $2\Msun$, and in the laminar model, the peak lifetime at $2\Msun$ is reduced to $\sim 20\Myr$.
    % However, the fast-wind model predicts estimated inner disk lifetimes longer than observations,  (Appendix~\ref{sec:inner_disk_lifetime}).
    However, the assumption of no wind mass loss in the fast-wind model seems difficult to reconcile with the high detection rate of [\ion{O}{1}]~$\lambda6300$ low-velocity component \citep[$\sim 80\%$;][]{2018_Nisini}, making it an unfavored PPD evolution scenario.
    The stellar-mass-dependent trends are consistent across a wide parameter space, as shown  by the population synthesis median in \fref{fig:multi_lifetimes} (cf. Appendix~\ref{sec:parameter_dependences}).
    Overall, varying the disk and wind parameters yields a wide range of possible lifetimes, especially near $1.5\Msun$. 
    We conclude that the average lifetime peaks at early A-type stars with $\tlife > 10\Myr$ for small-grain-depleted disks.

    We emphasize that the lifetimes derived here apply only to small-grain-depleted disks. 
    Figures~\ref{fig:t_life-Mstar} and \ref{fig:multi_lifetimes} neglect any short-lived disks that might be dispersed by FUV photoevaporation. 
    To quantify the general frequency of gas disks, it would be necessary to estimate PPDs' probability of becoming depleted in small grains.  This is an issue for future studies.

    We attribute the stellar mass dependence of $\tlife$ mostly to the variation with $M_*$ in the EUV and X-ray luminosities and thus the photoevaporation rates. 
    However, the variation of MHD effects with stellar mass may also play a role. 
    The hard X-rays from low-mass stars ($\lesssim 1\Msun$) can ionize the disk atmosphere to increase the local effective $\alpha$. 
    This enhances MHD winds' effects, shortening the disk lifetime. 
    The soft X-rays due to shocks originating from line-driven winds of early B- and O-type stars could have a similar effect and can also directly drive soft X-ray photoevaporation. % Brittain 2023
    Both these effects would strengthen the gas lifetime difference between the late intermediate-mass stars and the rest.

    Currently, no evidence supports the presence of companions in known gas-rich debris disks, but the possibility still remains that they could harbor unseen low-mass, close-in companions. In these scenarios, the EUV and X-ray emission may be dominated by these companions, boosting photoevaporation rates. Thus, to yield long-lived gas disks, late intermediate-mass stars should not have such companions. 
     %\citep{2022_Offner}

    % { Move this paragraph to summary(?)}
    % For a more comprehensive understanding, it is necessary to derive lifetimes with various model parameters, such as initial disk mass, viscous $\alpha$, wind torque, etc. 
    % We will address it in future studies. 
    % In any case, we expect the gas disk lifetimes to show qualitatively the same trend against the stellar mass as in \fref{fig:t_life-Mstar} since it is photoevaporation that sets the fate of the gas disk. 

    \section{Comparison with detection rate of gaseous debris disks}  \label{sec:observational_disks}
    % \cite{2017_Pericaud} performed a deep survey of young debris disks and found that the emission ratio of CO to the continuum ($S_\mathrm{CO}/F_\mathrm{cont}$) is 10-100 times higher for hybrid disks than the classical T Tauri stars, Herbig stars, weak-line T Tauri stars, and debris disks. 
    % The stars with a high $S_\mathrm{CO}/F_\mathrm{cont}$ tend to have a spectral type of A, a mass of $\sim 1.5$--$3\Msun$, a luminosity of $\sim 10$--$30\Lsun$, and an age of $\sim 5$--$30\Myr$. 
    % They concluded hybrid disks are the system where the dust has evolved more quickly than gas, being consistent with the assumption of our reduced photoevaporation model. 

    % The stellar types of the hybrid disks in \cite{2017_Pericaud} are all early A-type ranging in A3--A0. %with $T_\mathrm{eff} \gtrsim 8000\Kelvin$.
    % It agrees with our reduced photoevaporation model, where gas disks can survive longer, preferentially around early A-/late B-type stars. % earlier than A7 \citep[$\gtrsim 8000\Kelvin$][]{2013_Pecaut}. 
    % The consistency between the observations and our theory suggests that the gas in hybrid disks is indeed a protoplanetary remnant, as has been proposed in previous observational studies. 
    % For the first time, our model gives a physically natural reason for longevity and suggests a possible pathway for PPDs to become hybrid disks (\fref{fig:evolution_schematic}). 

    We now compare the modeled lifetimes from \secref{sec:lifetime_estimate} with the population of debris disks detected in the CO millimeter, [\ion{C}{2}]\,$157\um$, or [\ion{O}{1}]\,$63\um$ lines, focusing on the dependence on host star spectral type. 
    The detections come from Supplemental Table~1 of \cite{2018_Hughes}.
    Only 3 out of 152 sources listed in the table have ages of $< 10\Myr$, with all being A0~stars.
    % (HD~109573: $8 \Myr$, A0 star; HD~141569: $7 \Myr$, A0 star; HD~36546: $6 \Myr$, A0 star). 
    The age of HD~166191 (F star) is presented as $4 \Myr$, but we adopt an updated value of ${\sim} 10 \Myr$ \citep{2018_Potravnov, 2022_Su}.
    As they discussed, this sample combines studies with diverse sensitivity limits and selection criteria and thus is not well-suited for determining quantitative occurrence rates. 
    However, the spectral-type dependence resembles that in a more-uniform sampling by \citet{2017_Moor}.
    
\begin{figure*}
    % Database/MachineReadableTable/Hughes2018 : python3.9 hughes2018_table.py
    \centering
    \includegraphics[clip, width = \linewidth]{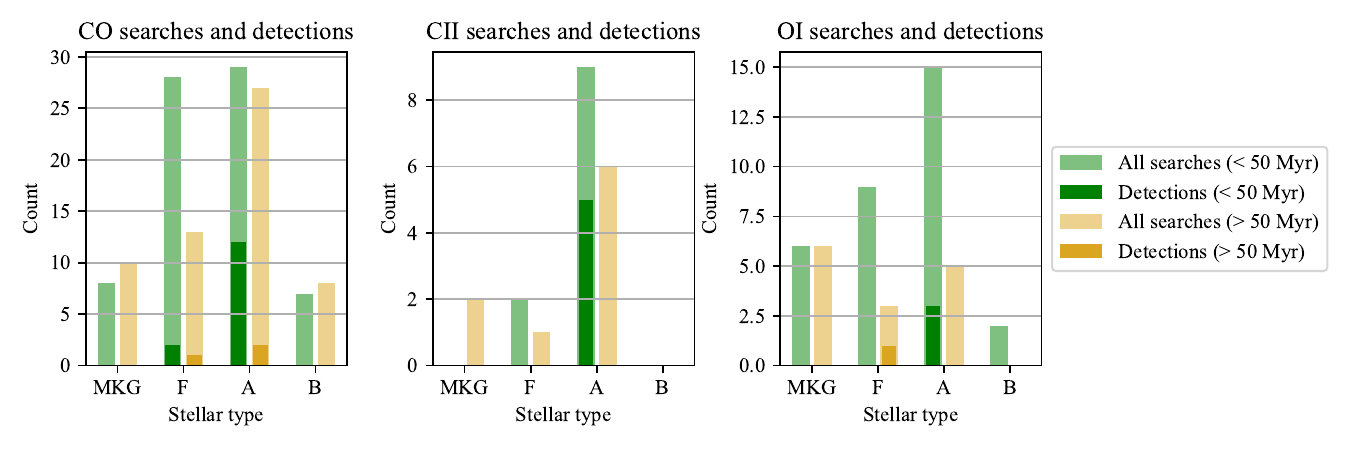}
    \caption{Detection rates of gaseous debris disks are highest among young A-type stars. 
    % All sources with gas detection are included regardless of the gas mass to maximize the samples. 
    The tracers are CO millimeter emission (left), \ion{C}{2}~157$\um$ (middle), and \ion{O}{1}~63$\um$ (right).
    The light-green histograms show the all-survey count and are superposed by dark-green histograms indicating the detection count, for sources younger than $50\Myr$. 
    The light-yellow and dark-yellow histograms are sources older than $50\Myr$. 
    The horizontal axis gives the stellar type. 
    Integrated counts are shown for M, K, and G stars. 
    All data are from \cite{2018_Hughes} Table~1.
    These trends are consistent with the $M_*$-dependence of our predicted lifetimes in \fref{fig:t_life-Mstar}. 
    }
    \label{fig:hughes_recompiled}
    \end{figure*}

    Keeping this limitation in mind, we replot the detections compiled by \cite{2018_Hughes} versus stellar type in \fref{fig:hughes_recompiled}. 
    We show not only the detections in CO but also those in \ion{C}{2} and \ion{O}{1}, since the disks may retain these species mixed with hydrogen even in the absence of CO.
    % Separating the detection frequencies for A- and B-type stars is important to test our hypothesized disk evolution model in the previous section, 
    % where we have proposed gas disks are relatively easy to survive around $2\Msun\lesssim M_* \lesssim 3\Msun$ compared to lower- and higher-mass stars.
    % \tref{tab:general_comparison} and 
    We note that nine of twelve young CO-detected sources have CO mass $M_\mathrm{CO} > 10^{-4}\Mearth$ \citep[cf.\ Table~A1 of][]{2020_Marino, 2017_Moor, 2019_Moor}. 
    The three sources below this threshold are likely better explained by secondary origins, based on the short photodissociation lifetimes for primordial CO at such low gas masses. 
    For F~stars, one of the two young CO-detected sources is above this mass threshold and is early F (F2/3V); the one below the threshold is late F (F5/6V).
    The decreasing trend of the incidence from A to FGK stars is likely real but not due to a sensitivity bias.
    %given the ALMA's ample sensitivity for CO detection at protoplanetary levels. 
    A comparison of debris disks with similar fractional luminosities \citep[Tables~A1 and A2 of ][]{2020_Marino} indicates that only 2 out of 9 sources exhibit gas (at levels $\lesssim 10^{-4} \Mearth$) among FGK stars, while 11 out of 17 A-type stars display gas presence (often exceeding $10^{-3} \Mearth$).

    \fref{fig:hughes_recompiled} shows several interesting consistencies with the modeled $\tlife$--$M_*$ relation. 
    The detection rate is highest for early A-type stars (more than half the stars with gas detected are early A-type, A0--A3). 
    Almost all the gas-rich debris disks with $M_\mathrm{CO}> 10^{-3}\Mearth$
    % namely
    % HD~156623 \citep[$3\e{-3}\Mearth$][]{2016_Lieman, 2019_Hales}
    % HD~32297 \citep[$1\e{-3}\Mearth$][]{2016_Greaves, 2017_Kral}, 
    % HD~131488 \citep[$9\e{-2}\Mearth$][]{2017_Moor}, 
    % HD~121617 \citep[$2\e{-2}\Mearth$][]{2017_Moor},
    % HD~131835 \citep[$4\e{-2}\Mearth$][]{2016_Lieman, 2015_Moor, 2019_Hales}, 
    % and HD~21997 \citep[$6\e{-2}\Mearth$][]{2013_Kospal}. 
    are hosted by early A-type stars. 
    The exception is HD~121191 \citep[$3\e{-3}\Mearth$, ][]{2017_Moor}, which is type~A5 but does not necessarily disagree with our model considering the range in $\tlife$ at $M_*\sim 1.5\Msun$. % with disk and wind parameters. 
    % For HD~32297 and HD~131835, C mass is measured and is also found to be massive, $M_\mathrm{C} = 3.5\e{-3}\Mearth$ \citep{2020_Cataldi} and $M_\mathrm{C} =  3\e{-3}\Mearth$ \citep{2019_Kral}, respectively. 
    Our model naturally explains younger ($< 50\Myr$) gaseous debris disks' higher numbers around A-type stars than around both lower-mass and O- and early B~stars. 
    % {Atypically low photodissociating fluxes of early A~stars would also be helping the long survival of CO.}
    The decreasing trend toward higher-mass stars could be due to both photoevaporation and intense CO photodissociation.
    For A~stars, the weak photodissociating fluxes would also be helping the long survival of CO.
    % To explain the decreasing trend toward higher-mass stars with secondary gas, gas removal processes, such as photodissociation and photoevaporation, would need to work more efficiently since higher-mass systems mean higher gas-production rates. 
    Overall, within the limitations of the sample, the consistencies with the model are remarkable. 
    % It proposes that the primordial-origin scenario should not be ruled out,  
    % but further detailed investigations are needed considering disk evolution from the protoplanetary state.

    In contrast, the gas detected in debris disks around late F-stars appears more compatible with secondary origins, given the short gas lifetime around those low-mass stars in our model. 
    % as the plausibility has been discussed in previous studies 
    % \citep{2016_Marino, 2016_Lieman, 2017_Kral}.
    Also, a few debris disks have multiple lines of evidence supporting a secondary origin, among them the A5--A6 star $\beta$~Pic
    \citep[e.g.,][]{2014_Dent, 2016_Greaves, 2023_Iwasaki, 2023_Cataldi}. 
    We thus consider primordial and secondary origins not mutually exclusive. 
    We stress that our hypothesized disk evolution does not necessarily rule out a secondary origin for some gaseous debris disks, even among A-type stars younger than $50\Myr$. 
    Determining the gas origins in individual sources is a challenge for the future. 
    However, the results we present here suggest the high overall frequency of gas-rich debris disks among A-type stars can come from the primordial origin.

    % The mean molecular weight of the secondary gas is about ten times heavier than the hydrogen-dominant primordial gas. 
    % It could hinder the disk gas evaporation from lasting for $> 10\Myr$. 
    % The narrow convective zone of F-type stars may also be helping the heavy gas disks to persist. 
    % If the F-type gas-rich debris disks are primordial in origin, it implies the applicability of the reduced photoevaporation model to F-type systems. 
    % It is worth being examined in future studies. 

     % The trend of higher gas detection rates around young intermediate-mass stars was previously noted by Lieman-Sifry et al. (2016), who surveyed 23 debris disk host stars in the 10-Myr-old Scorpion\UTF{2013}Centaurus region and detected strong CO emission from three of seven intermediate-mass stars but none from the 16 FGK stars.

    % \subsubsection{Connection to Herbig disks}
    % { TODO: Potential relation to Group~I Herbig disks. \citep{2015_Banzatti}, \citep{2018_Garufi}, \citep{2017_Garufi}.}

%------------------------- conclusions ---------------------
\section{Summary and Outlook}   \label{sec:conclusion}
The substantial amounts of gas found in some debris disks could either be remnants of the primordial protoplanetary disk, or originate in the secondary release of volatiles from bodies resembling comets or asteroids. 
% While there is a vast literature on secondary origins, less work has been done on the primordial-origin scenario. 
While the primordial-origin scenario was classically considered unlikely,
we here %examined the conditions under which 
{demonstrated} primordial gas can survive to the ages of the known gaseous debris disks, using semi-analytic models to derive lifetimes for the gas in disks that have been depleted in small grains ($\lesssim0.01\um$) so that FUV photoelectric heating is ineffective. 
The resulting lifetimes are longest around stars of $2\Msun \lesssim M_*  \lesssim 3\Msun$, whose evolution switches off their surface convection {at $1\text{--}10\Myr$} before the disk loses the bulk of its gas.
Without a surface convective zone, these stars lack surface magnetic activity and thus have low luminosities in the EUV and X-ray bands that drive the photoevaporation of the gas.

% In disks accreting solely under internal stresses, the gas survives $>10\Myr$ across all stellar spectral types. 
% In the wind-dominated disks, 
The gas lifetime \added{consistently} exceeds $10\Myr$ for the late intermediate-mass stars only. 
The shorter lifetimes \deleted{($< 10\Myr$)}of the gas disks around lower-mass stars ($\lesssim 1\Msun$) are due to strong EUV and X-rays from the chromosphere, transition region, and corona, while those around higher-mass stars ($\gtrsim 4\Msun$) are due to the high photospheric temperatures and resulting EUV luminosities. 

Our model predicts a higher frequency of gaseous debris disks around A- and late B-type stars compared to types both earlier and later. 
This is qualitatively consistent with the observed population, motivating further investigation of the primordial-origin scenario.
Longer gas disk lifetimes for A-stars could be a factor in the giant planet occurrence rate peaking at 1.7--$1.9\Msun$ \citep{2015_Reffert,2022_Wolthoff}, though more detailed modeling of the dissipating disks would be needed to assess the possibility of gas giant formation.  
% Learning how planet-forming disks are dispersed is indispensable to understanding planet formation.
Likewise, our results could relate to the increasing occurrence of transition disks with stellar mass \citep{2021_vanderMarel}. Further investigations are needed to evaluate whether these transition disks might serve as precursors to gaseous debris disks.

The primordial-origin picture allows the existence around late intermediate-mass stars of long-lived gas-rich disks with very weak dust emission, which we term ``phantom'' disks.  Finding such phantom disks would indicate that the abundant CO gas in at least some young debris disks is primordial.  Searches for gas-rich, dust-poor disks especially around A- and late B-stars aged $10$--$50\Myr$ could thus help determine the range of evolutionary paths followed by planet-forming disks. 
Note that the phantoms are different from the relic disks, purely viscous, dusty disks with central cavities larger than $100 \au$ surviving $>10 \Myr$ around Solar-type stars owing to very weak X-ray emission \citep{2011_Owen_b}.

{The fraction of PPDs achieving the small-grain-depleted state and the critical level of depletion required to inhibit FUV photoevaporation are open questions. 
Further hydrodynamics investigations are needed considering a range of stellar masses and FUV/X-ray luminosities.
We anticipate that PPDs with PAH/small-grain abundances below $\sim 0.1\text{--} 1\%$ of the interstellar level (see \secref{sec:lifetime_estimate}) may have reached this state and could be potential precursors of gas-rich debris disks around intermediate-mass stars.
PPDs with PAH/small-grain abundances {\it above} the critical level would have FUV-driven \ce{H2} winds extend to $\sim 10\text{--}100\au$ with temperatures exceeding a few hundred Kelvin and mass-loss rates of $\sim 10^{-9}\text{--}10^{-8}\Msun\yr^{-1}$ \citep{2018_Nakatani, 2018_Nakatanib, 2021_Komaki}.
Searching for these molecular winds could also help examine the depletion level in PPDs and very young ($<10\Myr$) hybrid disks (e.g., HD~141569), including hybrid disk candidates \citep{2023_Iglesias}, around weak X-ray emitters. 
}

{The point of} this work {is} %demonstrate 
that gas disks can persist beyond $10\Myr$, a necessary condition for the primordial-origin scenario. 
{A caveat is that our 0D approach cannot unambiguously predict the detailed}
% However, questions remain about the 
surface density evolution and the alignment with other observational characteristics of inner disks, such as accretion and wind mass-loss rates \citep{2018_Fang, 2023_Fang}.
% Therefore, conducting 1D simulations to follow the radial profile evolution accurately and comparing it with various observables is needed in future work. 
Future studies involve 1D simulations to accurately follow the radial profile evolution and make comparisons with those observables %compare them with those observables.
{
to evaluate the proposed evolutionary pathways.
Nevertheless, we expect the conclusion on the long lifetimes around intermediate-mass stars will remain consistent even in more advanced models, as the key determinant of outer gas disk lifetimes, which potentially manifest as gaseous debris disks in observations, is the photoevaporation rates set by the stellar emission rates. The speed of final-stage disk dispersal would not be significantly influenced by how the inner disks clear.
Thermochemical modeling would also help understand how the CO and C masses evolve in small-grain depleted disks and examine if the final values reproduce observations \citep{2023_Iwasaki,2023_Cataldi}.
For a broader perspective, it is worth investigating what fraction of PPDs in general can be long-lived considering environmental factors like external photoevaporation, tidal truncation, and late infall.
}

% To explain the relatively low detection frequency at B-type stars, the secondary-origin scenario may need mass-loss processes whose magnitude scales with the stellar mass.
% Further investigation is needed to accurately derive the gas disk lifetimes as a function of $M_*$ in large parameter space. 
% It is necessary to include the effects of, at least, dust dynamics, grain growth, stellar evolution, viscous accretion, and time-dependent mass-loss rates of MHD winds and photoevaporation. 

% \input{junk/summary_Nakatani21}

\acknowledgments %
We are grateful to the anonymous referees for their useful comments.
We thank Masanobu Kunitomo and Akimasa Kataoka for their practical and insightful comments on this study.
RN is supported by the Japan Society for the Promotion of Science (JSPS), Overseas Research Fellowship. 
SM is supported by a Royal Society University Research Fellowship (URF-R1-221669).
% The numerical computations were carried out on the Cray XC50 at the Center for Computational Astrophysics, National Astronomical Observatory of Japan.
This research was performed in part at the Jet Propulsion Laboratory, California Institute of Technology, under contract 80NM0018D0004 with the National Aeronautics and Space Administration and with the support of the NASA Exoplanets Research Program through grant 17-XRP17\_2-0081.
This work was strongly benefited from the Core2disk-III residential program of Institut Pascal at Universit\'e Paris-Saclay, with the support of the program ``Investissements d'avenir'' ANR-11-IDEX-0003-01. 

%\facility{JVLA}
\software{Numpy \citep{numpy}, Matplotlib \citep{matplotlib}, Astropy \citep{astropy:2013,astropy:2018,astropy:2022}, SciPy \citep{scipy}
}

\vspace*{50mm}

\bibliographystyle{aasjournal}
\bibliography{bibsamples}
% \bibliography{references}
%\bibliography{../../../template}

\setcounter{table}{0}
\renewcommand{\thetable}{\Alph{section}\arabic{table}}
\begin{appendix}

\section{The UV-switch Model}   \label{sec:uv_switch}
The evolution of an accretion disk redistributing its angular momentum via viscous friction can be described by 
\begin{equation}
    \frac{\partial \Sigma}{\partial t} 
    = \frac{3}{R} \frac{\partial}{\partial R} 
    R^{1/2} \frac{\partial}{\partial R} \nu \Sigma R^{1/2},
    \label{eq:surface_density_evolution}
\end{equation}
 \citep{1974_LindenbellPringle}.
If the disk evolves purely viscously with the viscosity profile of $\nu (R) \propto R$, 
the surface density profile $\Sigma(R, t)$ is analytically computed as  
\eq{
    \Sigma(R, t) = \frac{M_\mathrm{disk,0}}{2\pi R_1^2} 
                \braket{\frac{R}{R_1}}^{-1} 
                \braket{1 + \frac{t}{t_\nu}}^{-3/2} 
                \exp \braket{
                    -\frac{R}{R_1}
                    \braket{1+\frac{t}{t_\nu}}^{-1}
                    }
    \label{eq:sigmadot_ana}
}
where $M_\mathrm{disk,0}$ is the initial disk mass,
$R_1$ is the outer cut-off radius, 
and $t_\nu \equiv R_1^2/3\nu(R_1)$ is the viscous timescale at $R_1$. 
The nominal value of $t_\nu $ is 
\[
    \begin{split}
    t_\nu   &\sim 
        % 10 \Myr 
        0.1 \Myr
        \braket{\frac{\alpha}{10^{-3}}}^{-1}
        \braket{\frac{M_*}{1\Msun}}^{-1/2}
        \\
        &\quad \times 
        \braket{\frac{H(R_1; M_*, L_*)/R_1}{0.06}}^{-2}
        % \braket{\frac{R_1}{100\au}}^2
        \braket{\frac{R_1}{10\au}}^2
    \end{split}
    % In [72]: R1 = 100. * ms.const_au
    % In [73]: Omega = np.sqrt(ms.const_GMsun / R1**3)
    % In [74]: cs = np.sqrt(100.*(R1/ms.const_au)**-0.5 * ms.const_kB / (1.4/0.6)/ms.const_mH)
    % In [75]: nu = 1.e-3 * (cs/Omega)**2 * Omega
    % In [76]: t_nu = R1**2 / 3. / nu
    % In [77]: t_nu / ms.const_year/ 1.e6
    % Out[77]: 13.30973636134983
\]
with the so-called turbulent $\alpha$ viscosity, $\nu = \alpha \Omega H^2$ \citep{1973_ShakuraSunyaev}. 
We note that the scale height $H = \cs / \Omega$ depends on $M_*$ and $L_*$;
e.g., a passive disk has a temperature profile of 
\[
    \begin{split}
        T (R, L_*) &= T_\mathrm{eff} \braket{\frac{R}{R_*}}^{-1/2} \theta_\mathrm{inc}^{1/4}
        =  \braket{\frac{L_* \theta_\mathrm{inc}}{4 \pi \sigma_\mathrm{SB}R^2}}^{1/4}
        , 
        % In [61]: (ms.const_Lsun * 0.05 / (4 * np.pi * ms.const_Stefan * ms.const_au**2))**0.25
        % Out[61]: 186.26399744110142
    \end{split}
\]
where $\theta_\mathrm{inc}$ is the incident angle of irradiation,
and $\sigma_\mathrm{SB}$ is the Stefan-Boltzmann constant \citep{1987_KenyonHartmann, 1998_Hartmann, 2022_Pinilla}. 
The viscous $\alpha$ may also depend on $M_*$ as $\alpha \propto M_*$ so that the accretion rate reproduces the observational scaling profile $\dot{M}_\mathrm{acc} \propto M_*^2$ \citep[][also see discussions in \cite{2009_Gorti} and \cite{2021_Kunitomo}]{2003_Muzerolle, 2005_Muzerolle, 2004_Calvet, 2005_Mohanty, 2006_Natta, 2014_Alcala, 2017_Alcala, 2016_Manara, 2022_Manara}. % eq.20 in 2021_Kunitomo. 2004_Calvet for Intermediate-mass stars 2003_Muzerolle for low-mass stars. 2005_Muzerolle for substellar objects.
%\cite{2020_Manara} % accretors are revealed in > 5 Myr disks around low-and intermediate-mass mass stars (< 1.6 Msun) with accretion rates as younger sources . 
%\cite{2014_Yasui} % MIR lifetimes are shorter than NIR lifetimes for intermediate-mass stars and almost the same for low-mass stars.
% We also assume the viscous accretion flow timescale $t_\nu \propto M_*^{-1}$ so the steady-state accretion rate $\Mdot{acc,0} \propto M_\mathrm{disk, 0} t_\nu^{-1} \propto M_*^2$, as observed \citep[e.g.,][]{2004_Calvet, 2003_Muzerolle, 2005_Muzerolle, 2005_Mohanty, 2006_Natta, 2014_Alcala, 2017_Alcala, 2016_Manara, 2022_Manara}.

We can calculate the disk mass and mass accretion rate from \eqnref{eq:sigmadot_ana} as 
\begin{gather}
    % \dot{\Sigma}_\mathrm{acc} (R, t) = - \partial_t \Sigma = \braket{\frac{3}{2 }  - \frac{R/R_1}{ 1+ t/t_\nu}  } \frac{\Sigma}{t_\nu ( 1+ t/t_\nu)}\\
    %   sigmadot can have negative values at large radii because of spread.
    M_\mathrm{disk} (t) = \int_0 ^\infty \dd R \, 2\pi R \Sigma
                    =   M_\mathrm{disk,0} \braket{1+ \frac{t}{t_\nu}}^{-1/2}\\
    \dot{M}_\mathrm{acc} = - \dot{M}_\mathrm{disk}
                    = \frac{M_\mathrm{disk, 0}}{2 t_\nu}
    \braket{1 + \frac{t}{t_\nu}}^{-3/2}. 
    \label{eq:mdot_ana}
\end{gather}
% steady state accretion rate: $\dot{M}_\mathrm{acc} = 3\pi \Sigma $ 

When photoevaporation is taken into account in \eqnref{eq:surface_density_evolution}, 
the effect appears on the right-hand-side as a sink term
\[
    \frac{\partial \Sigma}{\partial t} 
    = \frac{3}{R} \frac{\partial}{\partial R} 
    R^{1/2} \frac{\partial}{\partial R} \nu \Sigma R^{1/2}
    - \dot{\Sigma}_\mathrm{ph}.
\]
The solution is the so-called UV-switch model \citep{2001_Clarke}, 
and its behavior is described as follows: 
(1) the disk evolves effectively purely viscously at the early stage. 
(2) After the accretion rate equals the photoevaporation rate, 
material supply from the outer part halts. 
(3) It quickly opens a cavity in the inner region. 
(4) Photoevaporation dominates mass loss afterward and determines the dispersal time of the remaining disk. 

The first transition point at (2) is when $\Mdot{acc}$ is reduced to $\mph$. 
This transition occurs at (using \eqnref{eq:mdot_ana})
\begin{equation}
    \begin{split}
    t_w & 
    = t_\nu \left[
    \braket{\frac{M_\mathrm{disk, 0}}{2t_\nu \mph}}^{2/3}   -1 \right]
    \approx t_\nu \braket{\frac{\Mdot{acc, 0}}{\mph}}^{2/3} 
    \\
    & \approx
    2.2 \Myr 
    \braket{\frac{\Mdot{acc,0}}{1\e{-6}\Msun\yr^{-1}}}^{2/3}
    \braket{\frac{\mph}{1\e{-8}\Msun \yr^{-1}}}^{-2/3}
    % In [105]: 0.1 * (1.e-6/1.e-8)**(2/3.)
    % Out[105]: 2.1544346900318834
    \end{split}
    \label{eq:switching_time}
\end{equation}
where $\Mdot{acc,0} \equiv M_\mathrm{disk,0}/2t_\nu$ is the accretion rate at $t = 0$. %% NOTE: Mdot_acc,0 must be larger than Mdot_ph
By this point, the disk mass has reduced to 
\begin{equation}
    \begin{split}
    & M_{\rm disk, 0}^\prime \equiv \Mdisk(t_w) 
     = M_\mathrm{disk, 0} \braket{1 + \frac{t_w}{t_\nu}}^{-1/2}
    = M_\mathrm{disk, 0}
    \braket{\frac{\Mdot{acc,0}}{\mph}}^{-1/3}\\
    & \approx 
    0.043 \Msun 
    \braket{\frac{M_\mathrm{disk,0}}{0.2\Msun}}
    \braket{\frac{\Mdot{acc,0}}{10^{-6}\Msun\yr^{-1}}}^{-1/3}
    \braket{\frac{\mph}{10^{-8}\Msun \yr^{-1}}}^{1/3}.
    \end{split}
    % In [106]: 0.1 * (1.e-6/1.e-8)**(-1/3.)
    % Out[106]: 0.02154434690031884
\end{equation}
Note that $M_\mathrm{disk, 0}$, $t_\nu$, and $\Mdot{acc,0}$ are not independent but are constrained by $M_\mathrm{disk,0} = 2 t_\nu \Mdot{acc,0}$.

Material supply from the outer disk stops at $t = t_w$, resulting in a fast draining of the inner region ($R\lesssim 1\au$) on the timescale of the viscous time at the critical radius $R\approx 1\au$, where a gap opens. 
Photoevaporation governs disk evolution afterward.
Note that for disks without MHD winds' mass loss, $t_w$ and the switching time of our model $t_0$ are identical.
The remaining outer disk disperses by photoevaporation on a timescale of the ``initial'' outer disk mass $M_{\rm disk, 0}^\prime$ over the photoevaporation rate,
\begin{equation}
    \begin{split}
    & \frac{M_{\rm disk,0}^\prime}{\mph}
     = 2 t_\nu 
    \braket{\frac{\Mdot{acc,0}}{\mph}}^{2/3}
    = 2 t_w \\
    & \approx 4.3 \Myr
    \braket{\frac{M_\mathrm{disk,0}}{0.2\Msun}}
    \braket{\frac{\Mdot{acc,0}}{10^{-6}\Msun\yr^{-1}}}^{-1/3}
    \braket{\frac{\mph}{10^{-8}\Msun \yr^{-1}}}^{-2/3}.
    \end{split}
    % In [108]: 0.1 * (1.e-6/1.e-8)**(-1/3.)/ 1.e-8 / 1.e6
    % Out[108]: 2.154434690031884
    \label{eq:evaporation_time}
\end{equation}
Hence, the {gas disk lifetime} is $\tlife = t_w + 2 t_w$ in cases of a constant $\mph$ before and after $t_w$.

Photoevaporation rates vary according to stellar evolution, where the luminosities and spectra change by orders of magnitude. 
% They also depend on the disk geometry; once a disk is inner-truncated, the inner rim is directly irradiated, which results in somewhat higher $\mph$ \citep{2006_Alexander_b}. 
% In such a case, the lifetime is shortened from $3t_0$, the classical picture of disk evolution. 
Also, if $\mph$ gets smaller at some point by, e.g., reduced photoelectric heating due to small-grain depletion, the lifetime is extended. 
This is the basic idea of our model \citep[and][]{2021_Nakatani} to explain a long-lived gas disk in the context of the primordial origin scenario.

\section{An analytic disk evolution model with MHD effects}     \label{sec:chambers}
The effects of magnetohydrodynamics (MHD) winds are essential in disk evolution. 
They can remove not only the mass but also the angular momentum of the disk. 
\cite{2019_Chambers} derived an approximate analytic solution for disk evolution with MHD wind effects. 
\eqnref{eq:surface_density_evolution} is modified to 
\begin{equation}
    \frac{\partial \Sigma}{\partial t} 
    = \frac{3}{R} \frac{\partial}{\partial R} 
    R^{1/2} \frac{\partial}{\partial R} \nu \Sigma R^{1/2}
    + \frac{1}{R} \frac{\partial}{\partial R} R v_w \Sigma 
    - \dot{\Sigma}_w,
    \label{eq:chambers2019}
\end{equation}
where $v_w$ is the inward velocity induced by the disk wind, 
and $\dot{\Sigma}_w$ is the surface mass-loss rate due to MHD winds. 
Using a reference radius $R_0$, the surface mass-loss rate is characterized by a constant parameter $K$ as 
\[
\begin{gathered}
    \dot{\Sigma}_w = - \frac{K f_w v_0 \Sigma}{R_0} \braket{\frac{R}{R_0}}^{-3/2}\\
    K = \frac{1}{2} \braket{\frac{R}{R_0}}^{-1/4} \braket{\frac{\Omega R}{v_\mathrm{esc} - \Omega R}}, 
\end{gathered}
\]
where $v_0$ is the inward velocity at $R_0$,
$f_w$ is the fraction of $v_0$ induced by MHD winds, 
$v_\mathrm{esc}$ is the tangential velocity of the winds,
and $\Omega$ is the Keplerian orbital frequency. 
The second equation is derived from the conservation of angular momentum. 

\cite{2019_Chambers} modified \eqnref{eq:chambers2019} to make it analytically solvable (Eq.(12) of the original paper) and found analytic solutions for the modified equation (Eq.(36)--(39) of the original paper). 
We omit to write the solutions explicitly here to avoid complexity and refer the readers to the original paper.
The point of this analytic solution is that the surface density and temperature evolutions are uniquely determined for a given set of model parameters $v_0$, $f_w$, $K$, the temperature at $R_0$: $T_0$, and the surface density at $R_0$: $\Sigma_0$. 

For example, $f_w = 0$ means purely viscous disk, and choosing a high $f_w (\leq 1)$ indicates a disk that accretes mainly magnetically. 
The parameter $K$ can be set to zero with a nonzero $f_w$. 
The underlying assumption, in this case, is negligible mass-loss rates due to MHD winds. 
Small $K$ cases are termed fast wind. 
On the other hand, using a high value for $K$ ($\leq 1$) means that a significant fraction of the accreting gas flows out of the disk before reaching the host star. 
This case is called a slow wind. 

The analytic formula of \cite{2019_Chambers} can overall well agree with the true solutions of \eqnref{eq:chambers2019} regardless of the approximation taken to make the equation analytically tractable. 
The exception is when the wind mass-loss rate is large; 
the analytic formula underestimates the surface density and outer spreading at the later stage. 
It also fails to predict the positive slope of $\Sigma$ created by the wind in the inner $\sim 1\au$ region. 
Nevertheless, the errors between the analytic formula and the true solution are small compared to the large uncertainties in the long-term evolution of disk evolution with the MHD effects. 
We, therefore, use the formula as a valid estimate for time-dependent disk mass evolution.

\section{Accretion-generated emissions}
\label{sec:accretion_generated}
   
We have omitted accretion-generated UV and X-ray emissions from our model. This extra radiation component can dominate the magnetic and photospheric components, leading to accelerated photoevaporation and reduced $\tlife$. 
We now examine if this omission affects the study's conclusions---specifically, the peak lifetime beyond $10\Myr$ at $M_* \approx 2\Msun$ (cf. Figures~\ref{fig:t_life-Mstar} and \ref{fig:multi_lifetimes}). 
Since the EUV models potentially account for the observed higher occurrence of gas-rich debris disks around A stars, our focus remains exclusively on the EUV models.

The exact portion of accretion luminosity ($L_\mathrm{acc} \approx GM_* \dot{M}_\mathrm{acc}/R_*$) contributing to EUV radiation is uncertain. 
For simplicity, we assume $\sim 4\%$ is processed into EUV emission, aligning with analogous estimation for FUV \citep{2008_Gorti} and X-ray \citep{2016_Hartmann}. 
This can be represented as:
\[
    \Phi_\mathrm{EUV, acc} \approx 2 \e{42} \sec^{-1}
    \braket{\frac{M_*}{1\Msun}} 
    \braket{\frac{\dot{M}_\mathrm{acc}}{10^{-8}\Msun\yr^{-1}}}
    \braket{\frac{R_*}{1\,R_\odot}}^{-1}, 
\]
Here, the average EUV photon energy is approximated to match that of $9000\Kelvin$ blackbody radiation, $14.5 \eV$. 
The total EUV emission rate is now $\Phi_\mathrm{EUV} = \Phi_\mathrm{EUV, ph} + \Phi_\mathrm{EUV, mag} + \Phi_\mathrm{EUV,acc}$.
In practice, accretion-generated EUV radiation may terminate once the disk develops an inner cavity, when photoevaporation and accretion rates balance at $\sim 1\au$ (the critical radius for photoevaporation). 
However, as the cavity-opening time is not determined unambiguously within the framework of our model, we let the accretion-generated EUV emission last until the disk dispersal; this scenario sets the lower limit for $\tlife$. 

\begin{figure}
    \centering
    \includegraphics[clip, width = \linewidth]{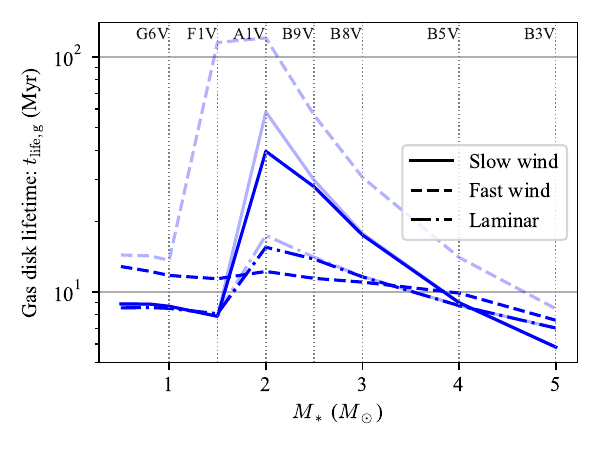}
    \caption{Derived $\tlife$ with accretion-generated EUV (blue). The lifetimes of the fiducial runs are also plotted for comparison (light blue).}
    \label{fig:phieuvacc}
    % python3.9 Dropbox/Liouhei_Ryohei/PythonScripts/HybridDisk_paperfigs.py life_multi_acc
\end{figure}
\fref{fig:phieuvacc} shows the resulting lifetimes. 
The accretion-generated EUV dominates $\Phi_\mathrm{EUV,mag}$ for the weak- and fast-wind disks especially at the early time ($\lesssim 1\Myr$), while laminar disks show minimal total emission rate enhancement.
Nevertheless, the photoevaporation rate remains orders of magnitude lower than the accretion and/or MHD wind mass-loss rates at the early stages, with no substantial impact on disk evolution. 
Except for fast-wind disks, the accretion-generated component becomes negligible as the accretion rate drops in $1\Myr \lesssim t \lesssim10\Myr$.
This occurs prior to the onset of the photoevaporation-dominant epoch at $t = t_0$, leading to minimal shortening of $\tlife$ for the slow-wind and laminar disks.

In contrast, fast-wind disks can maintain relatively high accretion rates ($\sim 10^{-10}$--$10^{-9}\Msun\yr^{-1}$) even at $t = 10\Myr$, with accretion-generated EUV dominating over (or comparable to) the other components throughout dispersal. 
This impacts $\tlife$ at $M_* = 1.5$--$2.5\Msun$, as $\Phi_\mathrm{EUV, acc}$ is significantly exceeds $\Phi_\mathrm{EUV, ph} + \Phi_\mathrm{EUV,mag}$.
However, the prevalence of jets and winds observed widely \citep[e.g.,][]{2022_Pascucci} suggests that the fast-wind scenario is unlikely to be common among PPDs. 
% Furthermore, this scenario likely may lead to overly long inner disk lifetimes (Appendix~\ref{sec:inner_disk_lifetime}). 
It is also difficult to align with the high detection rate of [\ion{O}{1}]~$\lambda6300$ low-velocity component \citep{2018_Nisini}.
If termination of accretion-generated emissions due to cavity opening is considered, the peak at $M_* \sim 2\Msun$ may remain even for fast-wind disks. % and/or UV/X-ray shielding by the inner disk winds is

Overall, incorporating the accretion-generated component can moderately reduce $\tlife$ and lessen the stellar-mass dependence for disks with no wind mass loss. 
Nonetheless, not all disks would evolve as fast-wind disks, supporting the highest plausibility of long-living ($>10\Myr$) gas disks around late intermediate-mass stars. 
Further investigation is needed to accurately determine $\tlife$, $\Phi_\mathrm{EUV,acc}$, and its termination time.

\section{Disk Parameter Dependences} \label{sec:parameter_dependences}
   
\begin{figure*}
    \centering
    \includegraphics[clip, width = \linewidth]{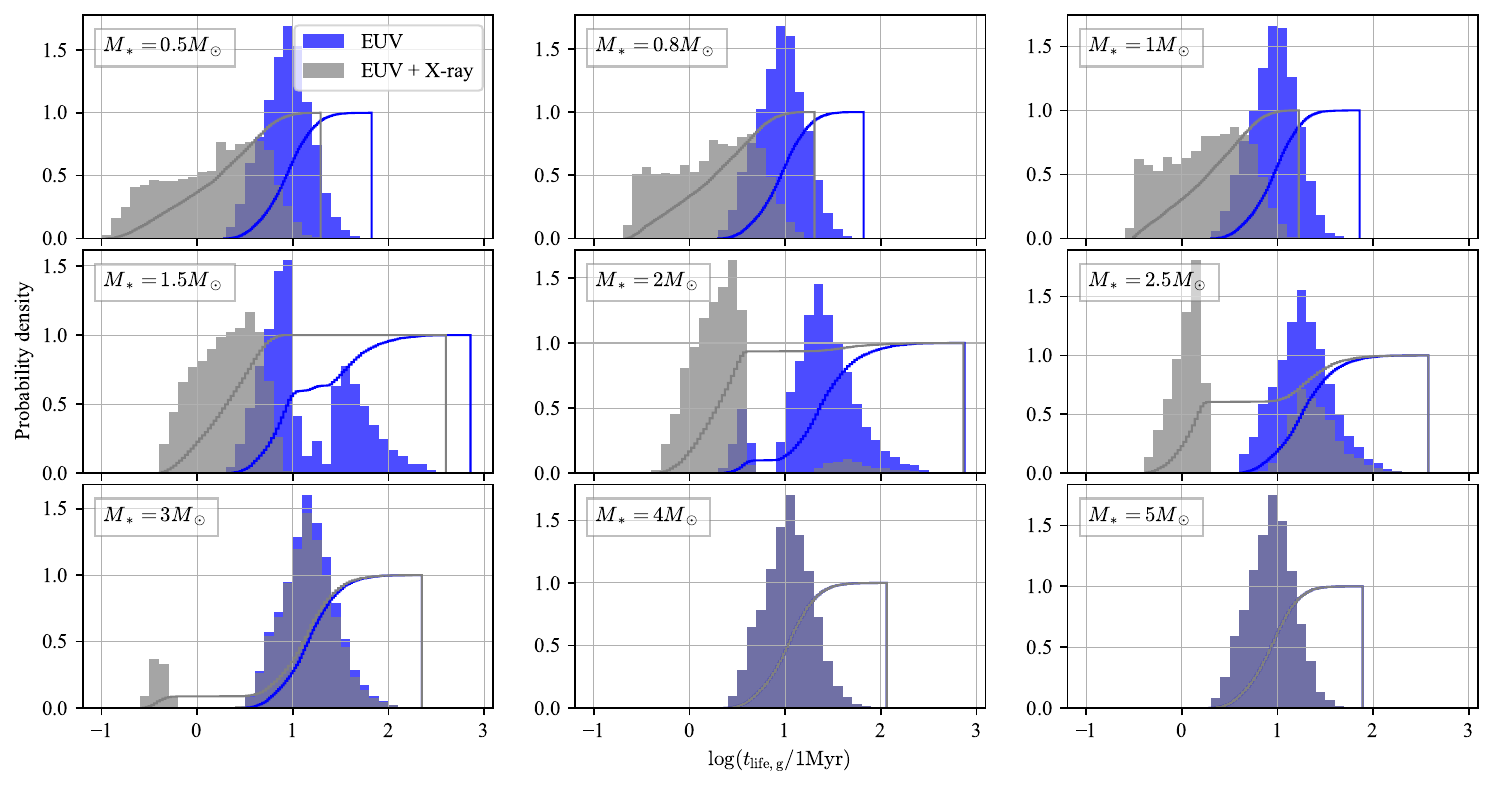}
    \caption{Resulting population. The blue and gray histograms are the probability densities for the EUV and EUV+X-ray models, respectively. Corresponding normalized cumulative probabilities are shown by the solid lines with the same colors. }
    \label{fig:population_synthesis}
    % /random2_generaterandomalpha5 : python3.9 ~/Dropbox/Liouhei_Ryohei/PythonScripts/HybridDisk_paperfigs.py life_rand2_all
\end{figure*}
We explore how $\tlife$ depends on various disk parameters, including the initial disk mass ($M_\mathrm{disk,0}$), disk mass scaling law ($M_\mathrm{disk,0} \propto M_*^\beta$), viscous $\alpha$, and initial disk cut-off radius ($r_\mathrm{exp}$).
These parameters are treated as Monte Carlo variables in population synthesis to span a wide parameter space. 

The initial disk mass is defined as $M_\mathrm{disk,0}/M_* = C_\mathrm{disk,0} (M_*/1\Msun)^{\beta-1}$, with our fiducial values being $C_\mathrm{disk,0} = 0.1$ and $\beta = 1$. 
Exploration covers a range of $0.03 \leq C_\mathrm{disk, 0} \leq 0.3$ for $C_\mathrm{disk,0}$ and $1 \leq \beta \leq 2 $, following observations \citep[e.g.,][]{2016_Pascucci, 2016_Ansdell, 2017_Ansdell}.  
As for $\alpha$, recent finding favor a low value at $10^{-4}\lesssim \alpha \lesssim 10^{-3}$ \citep{2022_Manara}, and therefore we vary it in $10^{-4}\leq \alpha \leq 2\e{-3}$, with the upper limit encompassing the slow-wind and fast-wind disks.
% The slow-wind, fast-wind, and laminar disk models of \citet{2019_Chambers} have the effective $\alpha$ of $\alpha_\mathrm{fiducial} \approx 0.002, 0.002, 3\e{-5}$, respectively. 

The fiducial $v_0$ values are $30\cm \sec^{-1}$ for slow-wind and fast-wind disks and $10\cm \sec^{-1}$ for the laminar disk. 
We explore $10\cm\sec^{-1} \leq v_0 \leq 30\cm \sec^{-1}$, as values beyond this range lead to unrealistic accretion rate variations. 
% Using $v_0$ of the order of, e.g., $\sim 1 \cm\sec^{-1}$ or $\sim 10^2 \cm\sec^{-1}$ results in too slow or fast accretion rate's decay with time and thus is likely unrealistic.
% These $\alpha$ and $v_0$ yield $f_w \gtrsim 0.8$, i.e., the disk accretion is dominated by magnetic braking. 
The wind mass-loss parameter $K$ is varied in $0 \leq K \leq 1$. The initial cut-off radius is tested in $10\au\leq r_\mathrm{exp} \leq 40\au$; the fiducial is $r_\mathrm{exp} = 15 \au$. 
For reference, \citet{2001_Clarke} and \citet{2020_Kunitomo} adopt $r_\mathrm{exp} = 10\au$ and $30\au$, respectively;
the time evolution of disk radius in the model by \citet{2022_Trapman} aligns with the characteristic radii of observed CO disks in Lupus and Upper Sco when $r_\mathrm{exp} \approx 20\au$ and $M_\mathrm{disk,0} = 0.1 \Msun$.

Monte Carlo variables assume uniform densities in linear space, except for $C_\mathrm{disk,0}$ and $\alpha$, which follow log-uniform sampling. %uniform lognormal distributions. 
We perform in total 90000 runs: 5000~parameter sets times 9 different stellar masses for each of the EUV and EUV+X-ray models.
Figure~\ref{fig:population_synthesis} illustrates the resulting probability densities and cumulative probabilities for $\tlife$, reflecting the trends seen in Figures~\ref{fig:t_life-Mstar} and \ref{fig:multi_lifetimes}. 

The lifetime is mostly unaffected by $\beta$ and $K$. 
For the EUV models and EUV+X-ray models with $M_*\geq 3\Msun$, $M_\mathrm{disk,0}$ has limited influence. 
However, for EUV+X-ray models with $M_*\leq 2.5\Msun$, smaller initial disk masses ($C_\mathrm{disk,0} \leq 0.1$) result in $\tlife \lesssim 5\Myr$. 
% The derived lifetime is not very sensitive to $M_\mathrm{disk,0}$ for the EUV models and EUV+X-ray models with $M_*\geq 3\Msun$. 
% For the EUV+X-ray models with $M_*\leq 2.5\Msun$, almost all disks with small initial disk masses ($C_\mathrm{disk,0} \leq 0.1$) have $\tlife \lesssim 5\Myr$. 

Increasing $r_\mathrm{exp}$ moderately extends $\tlife$. 
In the EUV models, $\tlife$ tends to exceed $10\Myr$ for the runs with $M_* \leq 1.5\Msun$ and $r_\mathrm{exp}(> 25\au)$.
Conversely, the most runs with $M_* = 2$--$3\Msun$ and $\tlife < 10\Myr$ have $r_\mathrm{exp} \lesssim 20\au$. 
% The lifetimes are relatively sensitive to $r_\mathrm{exp}$ for the EUV models with $M_* \leq 1.5\Msun$. 
% Disks with larger $r_\mathrm{exp} (> 25\au)$ account for the most of long-living disks ($\tlife > 10\Myr$).
When $r_\mathrm{exp}$ is expanded while keeping $M_\mathrm{disk,0}$ constant, it decreases the initial surface density in the inner part, leading to greater mass distribution in the outer disk.
This leads to slower mass loss via accretion and MHD winds. 

Viscous $\alpha$ and $v_0$ also have a moderate impact on the EUV models, with smaller $\alpha$ and higher $v_0$ causing shorter lifetimes due to enhanced MHD-driven mass loss and accretion. 
The longest-$\tlife$ population ($> 10^2\Myr$) at $M_* = 1.5$--$3\Msun$ has $\alpha > 10^{-3}$ and $v_0 < 15\cm \sec^{-1}$, indicating a slow decay of the accretion rate. 
Unrealistically long lifetimes ($\sim 10^3 \yr$) can occur when $\alpha$ is increased to $\sim 10^{-2}$.
Hence, when the disk exhibits a characteristic of a viscous disk relatively strongly, $\tlife$ increases significantly.

% Similarly, the population is not strongly dependent on $\alpha$. There is a trend that a smaller $\alpha$ results in a shorter lifetime since the more magnetized, the more quickly the disk mass reduces through disk wind and magnetic braking. 
% Increasing $\alpha$ to, e.g., $\sim 10^{-2}$ results in unrealistically long lifetimes ($\tlife \sim 10^2$--$10^3\Myr$) especially at $M_* < 3\Msun$. 
% Thus, a high $\alpha$ (purely viscous disks) are not preferred, which agrees with recent observations \citep[e.g.,][]{2016_Pinte, 2017_Rafikov, 2022_Manara}. 
% The population is not very sensitive to $v_0$, with a higher $v_0$ resulting in slightly shorter lifetimes, and is almost independent of $K$. 

In summary, higher $\alpha$ and larger $r_\mathrm{exp}$ can extend $\tlife$ for specific cases. 
The other parameters exert minor impacts. 
Regardless of the parameter dependences, overall trends remain consistent: 
Late intermediate-mass stars in the EUV models consistently exhibit $\tlife > 10\Myr$, peaking at $M_* = 2\Msun$ and declining for lower and higher masses. 
While the fraction of long-living ($\tlife > 10\Myr$) disks for $M_* \leq 1\Msun$, most of which have $r_\mathrm{exp} > 25 \au$, is apparently higher than observations (\fref{fig:hughes_recompiled}), the actual CO disk lifetime is shorter than $\tlife$. This is because complete CO photodissociation occurs before disk dissipation, when $\Mdisk$ reduces to a certain threshold mass.
In addition, X-ray photoevaporation likely causes mass loss at somewhat much smaller rates than \eqnref{eq:mdot_owen}, implying that the the lifetimes of the EUV models should be interpreted as upper limits for low-mass stars.

Thermochemistry modeling is needed to interpret these results fully, along with the potentially crucial influence of atypically weak stellar photodissociating fluxes around A~stars. 
It is important to highlight that our findings primarily demonstrate the plausibility of gas disk survival beyond $> 10\Myr$, a necessary condition for supporting the primordial-origin scenario. 
Additionally, our model qualitatively explains the observed incidence of gaseous debris disks versus stellar mass.  
% \input{junk/parameter_dependence}

% None 0.5 3609 8.621009818927359 1.748239950361057
% None 0.8 3609 9.06551008002394 1.748109539461287
% None 1.0 3609 9.110348125299385 1.728182070851564
% None 1.5 3609 14.212984952693166 2.955038550939682
% None 2.0 3609 22.940022113136035 2.455895553625839
% None 2.5 3609 18.789256998883825 1.9808234691338427
% None 3.0 3609 14.908185486119825 1.885163512396364
% None 4.0 3609 10.781379648943892 1.7813723339300604
% None 5.0 3609 8.588880333808861 1.7278839086781954

% None 0.5 5122 8.61847521931534 1.7446756641908527
% None 0.8 5122 9.063143060748365 1.7449084856656245
% None 1.0 5122 9.108422955400915 1.7253340382095343
% None 1.5 5122 14.217827974098277 2.9618438642838325
% None 2.0 5122 22.98204408126579 2.4537819513176977
% None 2.5 5122 18.809670077627672 1.9794734896129647
% None 3.0 5122 14.91699661462661 1.8836925239247608
% None 4.0 5122 10.780685684282366 1.7800765279588766
% None 5.0 5122 8.584926782579327 1.7271222494336274
% Owen+12 0.5 5122 1.4765413115303552 3.141568091732717
% Owen+12 0.8 5122 1.6892359605433724 2.8157752256853734
% Owen+12 1.0 5122 1.7834330314912228 2.569195625578189
% Owen+12 1.5 5122 1.9891806598087456 2.0527874525450427
% Owen+12 2.0 5122 2.214375244521114 2.6909525008208153
% Owen+12 2.5 5122 3.7590419494517477 4.898141605329055
% Owen+12 3.0 5122 10.612690844347485 3.263978571281137
% Owen+12 4.0 5122 10.780685684282366 1.7800765279588766
% Owen+12 5.0 5122 8.584926782579327 1.7271222494336274

\section{X-ray Luminosity Spread}
\label{sec:X-ray_spread}

Observed X-ray luminosities are known to have a large spread \citep[e.g.,][]{2007_Gudel}.
Accounting for the variability in X-ray luminosity would introduce scattering in the derived lifetimes. 
In the EUV+X-ray models, amplifying X-ray luminosity results in shorter lifetimes while reducing it aligns the lifetimes with those of EUV models. 
Within the EUV models, the spread in luminosity could influence lifetimes for $M_* < 1\Msun$, as their lifetimes are linked to $\Phi_{\mathrm{EUV, mag}}$. 
However, combining Eqs.~(\ref{eq:mdot_euv}) and (\ref{eq:Phi_EUV_mag}), the scaling of EUV photoevaporation rate with $L_X^{0.33}$ places a relatively restrained impact on lifetimes. 
Furthermore, the adopted X-ray luminosity encompasses representative observational values (Fig.6 of \citet{2021_Kunitomo}). 
It suggests that derived lifetimes remain relatively stable even when incorporating luminosity spread with a plausible distribution of $L_\mathrm{X}$.
Consequently, the inclusion of luminosity dispersion is unlikely to significantly alter the core finding of this study.

Taking into account the spread of $L_\mathrm{X}$ along with its time evolution would require solving stellar evolution across a wider parameter space than covered by \citet{2021_Kunitomo}.

\end{appendix}

\end{document}